# Toward a new approach for massive LiDAR data processing

V-H. Cao[#1], K-X. Chu[#2], N-A. Le-Khac[#3], M-T. Kechadi[#3], D. Laefer[*3], L. Truong-Hong[*3]

[#]School of Computer Science & Informatics, University College Dublin
Belfield, Dublin 4, Ireland
[1]van-hung.cao@ucdconnect.ie
[2]chuxuankhoi@gmail.com

[*] School Of Civil, Structural & Environment Engineering, University College Dublin
Belfield, Dublin 4, Ireland
[3]{an.lekhac,tahar.kechadi,debra.laefer,linh.truonghong}@ucd.ie

*Abstract*— **Laser scanning (also known as Light Detection And Ranging) has been widely applied in various application. As part of that, aerial laser scanning (ALS) has been used to collect topographic data points for a large area, which triggers to million points to be acquired. Furthermore, today, with integrating full wareform (FWF) technology during ALS data acquisition, all return information of laser pulse is stored. Thus, ALS data are to be massive and complexity since the FWF of each laser pulse can be stored up to 256 samples and density of ALS data is also increasing significantly. Processing LiDAR data demands heavy operations and the traditional approaches require significant hardware and running time. On the other hand, researchers have recently proposed parallel approaches for analysing LiDAR data. These approaches are normally based on parallel architecture of target systems such as multi-core processors, GPU, etc. However, there is still missing efficient approaches/tools supporting the analysis of LiDAR data due to the lack of a deep study on both library tools and algorithms used in processing this data. In this paper, we present a comparative study of software libraries and algorithms to optimise the processing of LiDAR data. We also propose new method to improve this process with experiments on large LiDAR data. Finally, we discuss on a parallel solution of our approach where we integrate parallel computing in processing LiDAR data.**

## I. Introduction

Air-borne Light Detection And Ranging (LiDAR) has been used to acquire three dimensional (3D) topographic data points of object's surfaces of a large coverage area. LiDAR data has been widely used in a large range of applications, such as forestry management [1], 3D city modelling [2], road detection [3] disaster management [4] and computational modelling [5]. Recently, an aerial laser scanning (ALS) has a scan rate of 1MHz and Full Waveform Digitizer (FWD) collection at up 120 kHz [6], where ALS data consist hundreds of millions of 3D point clouds associated with waveform data of laser pulses. For example, with specific designated scanning plan, ALS data of 1 km$^2$ Dublin city center is of 225 million points with around 5.9 GB in size, acquired by the Urban Modelling Group at the University College Dublin [7]. In addition, the national wide LiDAR data in Netherland has around 640 billion points, where each block 1000x1250m of ASCII xyz data (each LiDAR point stored with x, y and z coordinates) is approximately 0.5 TB in size [8, 9]. Raw ALS data recorded are needed to process in order to obtain real 3D point clouds. Unlikely traditional ALS data acquisition recorded four or more returns per transmitted pulse, the modern ALS with FWF data contain additional up to 256 samples for each return pulse. In this context, volume and complexity of input ALS data have been increased dramatically.

The huge volumes and complexity of ALS data are to be great challenges for data processing as the limitation of the computing hardware. With conventional sequence algorithms, ALS data processing is to be time consuming because the processing is computationally intensive and iterative. For example, the total execution time on 2-way-Quad-core computer was approximate 2 hours to generate Delaunay triangulation of 0.883 billion LiDAR points occupied 16.4 GB in size [10]. Moreover, for segmenting 105 millions of mobile laser scanning points, the shape-based segmentation method takes about 1 hour, where the experiment was perform on a machine with 8 GB RAM and an Intel Core i3 with a speed clock of CPU by 3.07 GHz [11]. Thus, development of alternative solutions is urgently needed in practical applications. For example, after natural disaster, digital elevation model (DEM) is quickly requested for damage estimation. Various optimization techniques and algorithms have been proposed to improve performance of LiDAR data processing [12,13,14]. Ones of which, parallel processing is to be a potential ALS data processing solution [15,16]. However, these approaches are normally designed with regard to the parallel architecture of target systems such as multi-core processors, GPU, etc. There is still missing efficient approaches/tools supporting the analysis of LiDAR data due to the lack of a deep study on both library tools and sequential algorithms used in processing this data. Thus, this paper investigated current efficient techniques and proposed a new strategy in ALS data processing. Eventually, in this paper, we firstly present a comparative study of software libraries and algorithms to optimise the processing of LiDAR data. Next, we propose new method to improve this process with experiments on large LiDAR data. Finally, we discuss on a parallel solution of our approach where we integrate parallel/cloud computing in processing LiDAR data.

The remainder of the paper is arranged as follow. The background is introduced in following section where we review different approaches of processing LiDAR data. Next, we conduct comparative studies of existing libraries and methods to determine the key issues that affect the performance of LiDAR processing. Following, the new strategy for ALS data processing is introduced. We also discuss on the parallel approach for our strategy. Finally, conclusions and further work are drawn.

## II. BACKGROUND

Our long term goal is to develop efficient algorithm based on high performance computational (HPC) resources for classifying ALS data points into separate categories and for extracting the point cloud of separate objects. In the light of these aim, this section investigates commonly existing approaches used to solve these problems. Many algorithms have been proposed to automatically extract ground points and non-ground points from ALS data points, which are based on assumption about a structure of bare-Earth points in a local neighbourhood [17]. These methods can be divided four distinct groups: (1) slope-based: when the slope or height different between two points exceeds the certain threshold, the highest point is classified to belong to the object [18, 19]; (2) cluster/segmentation-based: the method is based on an observation is that the objects always have distinct edges to the bare-Earth, and any point within the closed boundaries of the cluster or segment known as the object is assumed as a part of the bare-Earth and other points of the cluster/segment are of the object [20]; (3) surface based: the points are within the buffer zone of the surface are assumed as the ground points, where the surface is iteratively determined from the points and the vertical distance of the points to the surface as the weight function [21-23]; and (4) morphological filter: the method is based on a series of opening operations to eliminate non-ground points [24].

Furthermore, the segmentation process, is to partition 3D ALS point clouds into subsets satisfying certain criteria [25], can be roughly classified as model fitting-based methods [26, 27], region growing-based methods [20, 28] and clustering feature based methods [29, 30]. For the last two segmentation categories, the key parameters of those algorithms involve a normal vector, a distance between a point to a fitting plane, a curvature of each point and a slope computed from a given point and its neighbourhood. Thus, the quality of the segmentation depends on selecting the neighbouring for computing these features, whereas the nearest neighbour search (NNS) is an important aspect in the algorithm. That is because the NNS is dominant execute time in computing variables for discriminant function of the classification and segmentation of ALS data. In summary, irrespective classification or segmentation process, since these methods are based on a local neighbourhood, the nearest neighbour search plays out an important role in controlling the performance of the algorithm. However, conventional sequence methods are to be computational overhead with massive ALS point clouds. Development of an efficient algorithm is therefore necessarily for improving the performance. The following section reviews parallel processing in building a tree structure supporting for NNS procedure.

## III. COMPARATIVE STUDY

### A. Comparative state-of-the-art approaches in HPC for LiDAR processing

In this part, we will take a look at some current approaches in HPC for LiDAR data processing. It will convey a snapshot of the state-of-the-art in this field and offer a viewpoint of the potential as well as rising challenges of applying HPC to LiDAR processing. In particular, the HPC-based paradigms in this part comprise cloud computing environments, PC Cluster, field programmable gate array (FPGAs), hardware systems such as multi-core CPU architecture, graphic processing units (GPUs), and general-purpose computing on graphics processing units (GPGPUs).

Xuefeng Guan and Huayi Wu (2010) [43] leveraged the power of multi-core platform to deal with massive geospatial data. They divided raw data into overlapped blocks and inputted concurrently these blocks on parallel pipelines. Multi-thread was used to exploit the full power of a multi-core processor. However, there are several drawbacks of multi-processor pipeline architecture which are listed by Duoduo Liao and Simon Y. Berkovich such as the amount of memories of processors duplicated (due to overlapped data), bus traffic problem. They proposed a new multi-core pipelined architecture [44] based on crossbar switching. Comparison between the new architecture and conventional multi-core architecture show that the new one gives much better performance than the old. Moreover, the new architecture significantly overcomes all the limitations of multi-processor pipeline.

Parallel processing methods using GPUs & GPGPUs also have been introduced to speed up computation recently. Hu et al. proposed a simple scan-line-based algorithm using parallel computing [45]. Authors propose a scan-line segmentation (SLS) algorithm to classify ground and unground object based on the calculation of slop and elevation. Using GPU's thread blocks to calculate parallel scan lines, each thread block of GPU process one scan line.

FPGA-based computing could offer on-board real-time processing [46]. Multi-level parallelism inherent in algorithms used for LiDAR processing could be exploited to speed up the procedure using High-Performance Embedded Computing (HPEC) systems featuring FPGAs. FPGA-based computing could be fast and fully reconfigurable now, but the developing time price are still high. Thus, it will not really appropriate for regular developer [45].

### B. Software study

As mentioned above, software is an important part in processing LiDAR data. It is not only used to visualise the experimental results in the way that everyone can understand but also provide the numeric data (such as processing time, statistical results or other useful information for users in analysing LiDAR data) quickly and in convenient forms. However, software itself contains the problems that impact strongly to the users feeling. One of example is the duration of calculation. If the duration is too long and the application is lack of solutions to notice the progress, users may be confused to decide to stop the application or continue waiting. Another example is the performance of painting data on the screen. If the application is lack of solution to display smoothly when interacting, the movements may not be as the expected from users and they may reject the application before it has opportunity to show the advantages.

We test first of all the loading LiDAR data from file (using laspy library [32]) and buffering vertices to display (using *glBufferData* of OpenGL).

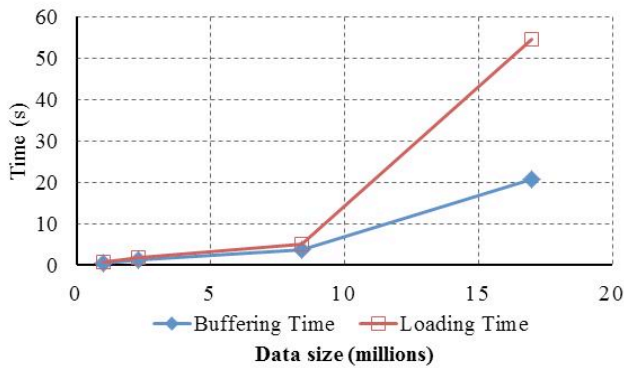

Fig.1. Loading time when using **laspy** & Buffering time when using **OpenGL**

Looking at Fig.1, we can see that when data size is less than 10 million points the loading time is accepted but when the size of the data is greater than 17 million points, loading time is significantly increased. Note that in the real-world application, data size in LiDAR processing is not just 17 million, it is usually from tens to hundreds million points.

Moreover, when working with huge data (hundreds of MB to GB), sometimes, RAM (which secures the fastest accessing speed for processing) cannot provide enough space to store and calculate; thus, the performance of the system falls down strictly (users have to wait in minutes to see the results). For example, when loading about 17 million points with a 4G RAM computer (running with the tasks of Windows 7 and Mozilla Firefox), the physical memory is overflowed. Of course, the problem does not appear with all computers, but in general it will impact to a part of users.

In this section, some aspects of software development are discussed to provide useful information for selecting or improving a performance of the software in processing LiDAR data.

1) *Python language*: Python is a very-high-level dynamic object-oriented programming language. It is chosen as the language to review and implement in the project because of the following reasons: (i) Python is dynamic, (ii) Broad standard library and portable and (iii) Huge community of users and developers [33].

2) *Libraries:* as mentioned above, many libraries developed by Python community are useful in managing and processing data. NumPy [34] (or its extension SciPy[35]): a powerful library to make calculations with multi-dimensional array objects. In fact, NumPy provides surprising manipulations such as broadcasting functions or memory management. However, it seems very bad in iterating over the data in arrays as shown in Fig. 3 where the computational time is linear with the number of data points.

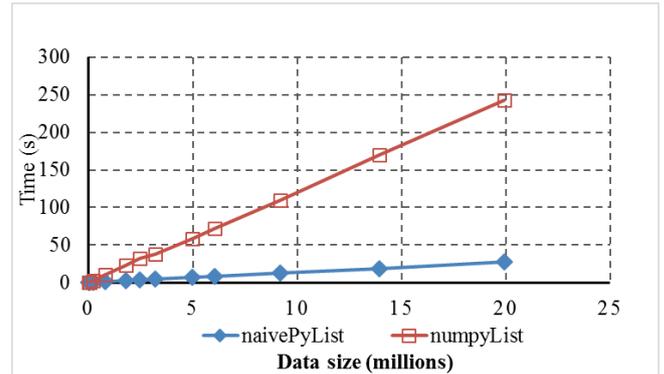

Fig.2. Comparison between looping with Python list and NumPy array

Beside the specific packages, SciPy provides a very interesting package named weave, which allows developer to embed C/C++ code to improve the performance [36]. Although the package needs extensive investigation to prove, it is considerable because of the proved efficiency and the flexibility of C/C++.

Built-in library: Python itself is updated and improved day by day, so, in almost time, using built-in functions is the most convenient and efficient way to develop. One example was mentioned in section that introduces about NumPy.

IV. TOWARD A NEW APPROACH FOR PROCESSING VERY LARGE LiDAR DATA

In this section, we propose a new approach to improve ALS data processing. Eventually, in our approach we look at two levels of optimisation: loading data and building data structure.

A. Loading data:

In order to reduce the time of processing, there are some strategies considered and investigate. However, the results are not as expected. Following is the strategies in specific:

Vectorising: NumPy also provides some functions to vectorise a sequence. However, NumPy document [37] and some discussions [38] pointed out that the performance would not be improved. The strategy requires the actions in deeper layers by using other languages such as C and Assembler.

Multi-threading: other strategy is often used in modern software development based on its ease of implementation is dividing the task to many subtasks and performing them on many parallel threads.

B. *Application of data structure:*

In LiDAR processing and visualization, searching neighbourhood of a given points is dominated a task. Data structures are necessarily to use for a massive data in order to provide efficient searching process, whereas *kd-tree* or *octree* are often used. Indeed, based on our comparative study, in order to improve the performance of LiDAR processing, this section is investigated appropriate data structures for massive LiDAR data. This investigation will provide useful information to establish high performance computing (HPC) strategy in LiDAR processing.

Although various data structures (i.e. kd-tree, octree or R-tree) have been widely used a *kd-tree* [39] a binary tree structure is primarily investigated in this section as its simple form Each non-leaf node corresponds an axis-aligned rectangular cuboid and its children split up the volume to form smaller cuboids through a splitting hyper-plane [48]. The constructing procedure of *kd-tree* has $O(kNlogN)$ space complexity [31]. A *kd-tree* can be used to accelerate *k-nearest neighbour (kNN)* queries [48] by using ball-rectangle intersection tests. With a given point *p*, a ball centred at *p* passes through the current *kth-nearest candidate* with the average complexity $O(log\ n)$.

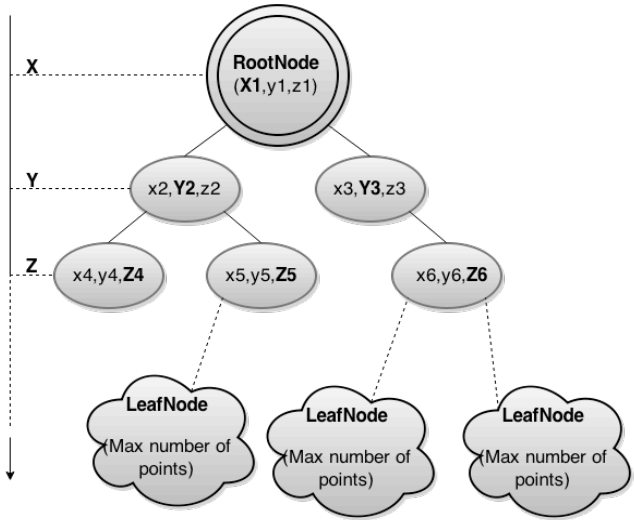

Fig.3. New proposed 3-d tree

In fact, the use of kd-tree affect to the performance of data processing, whereas involves time for building kd-tree, searching kNN of the given points in the data set and memory consumption. Before moving the parallel solution, a new approach was proposed to construct *kd-tree* with various sizes of the leafnode to investigate executing time, which contains a maximum number of points. Since raw LiDAR data often contain x-, y- and z- coordinates associated attributes (i.e. intensity value and red-green-blue values), we decided to build a 3d-tree to present all points in the dataset based on [39]. Each Leafnode is a bucket, which contains a maximum number of points. Fig.3 shows the *3d-tree* built by our new approach.

The new algorithm of constructing *3d-tree* is described as below:

*Algorithm 1 Optimal kdContruct*
1: **procedure** kdConstruct (*trainingSet*)
2: **if** *trainingSet.size()* <= Leafsize **then**
3: **return** *kdtree.leafnode*     // Returns a kdTree
4: **else**
5: (*s, val*) ← chooseSplit(*trainingSet*) // s is splitting dimension, chooseSplit function based on sliding midpoint rudes
6: *trainLeft* ← {*x* ∈ *trainingSet* : $x_s$ < *val* }
7: *trainRight* ← {*x* ∈ *trainingSet* : $x_s$ ≥ *val* }
8: *kdLeft* ← kdConstruct(*trainLeft*)
9: *kdRight* ← kdConstruct(*trainRight*)
10: **return** *kdtree*(*s, val, kdLeft, kdRight*)
11: **end procedure**

There are several split rule such as Standard Split Rule [31], Midpoint Split Rule [40]. However, in our algorithm we used Sliding Midpoint Split Rule [47] to ensure that the cells do not all become long and thin.

When we NN query *kNN* of a given point *p*, the procedure normally searches close-points in the tree and if some points stored in a Leafnode, it will use brute-force algorithm to find theses. Details of the searching procedure can be found in [31]. The algorithm of brute-force can be shown as follows:

*Algorithm 2 Brute-force*
1: c ← first(P) // generate a first candidate solution for P
2: **while** c < > Λ do
3: **if** valid(P,c) **then** output(P, c) // check whether candidate c is a solution for P then return output c
4: c ← next(P,c) // generate the next candidate for P after the current one c
5: **end while**

The dataset used in experimental tests include approximately 18 million points. The testing platform is Intel Core i7-3517U 1.9 GHz Processor, 2 GB DDR3 RAM, 256 GB Solid State Drive, Windows 8.1. We ran system to build *3d-tree* with Leafnode size = 1k, 5k, 10k, 50k, 100k, 200k, 500k, 1,000k points and kNN searching with 50 *kNN* was used to investigate running time for searching neighbourhood by using a specific *kd*-tree, whereas 10 given points were used to determine average processing time for searching the neighbourhood of one data point. The constructing tree time and searching time were shown in Fig. 6.

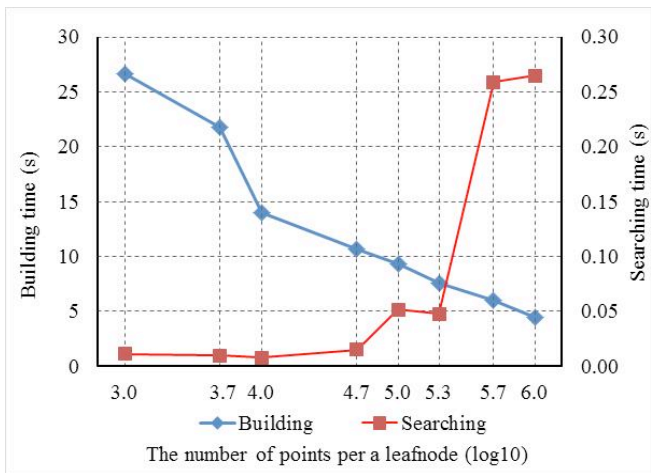

Fig.4 The average processing time of difference cases of Leafnode

As shown in Fig.4, processing time of constructing *kd-tree* decreased dramatically when the leafnode size of the tree increased. Obviously, once increasing Leafnode size, the shorter tree built reduced the constructing time and save the memory to store tree and boost CPU processing time. Furthermore, observing executing time of searching *kNN* of a given point, for *kd-tree* with a leafnode size in range from 1k to 10k, it is seemly constant and to be increasing for the rest of *kd-tree*. Particularly, when the *kd-tree* with the leafnode size by 1,000k points, the searching time increased approximately 34 times compared to the *kd-tree* with the leafnode size by 10k points. Thus, although the constructing time of the *kd-tree* is still very small compared the searching time of *kNN* of all data points in the data set, the *kd-tree* with the leafnode size by 10k points is arguably recommended.

Indeed, we can also improve our new approach of building *kd-tree* from *Algorithm 1* by using parallel approach for both multi-core/multi-processor platforms and distributed platforms. Recently, we proposed an effective network topology *TreeP* [41] for deploying parallel computational tasks on centralised and distributed platforms.

Eventually, the *TreeP* structure is similar to a *B+Tree* [42]. However, unlike *B+Tree*, its leaves' nodes (level 0) can also be part of any other levels. The higher level nodes act as a fabric of a virtual interconnection network, and are called virtual nodes. Another main difference between *TreeP* and *B+Tree* is that the nodes within the same level are connected by a bus topology, hence avoiding unnecessary communication through other levels. One of the most interesting features of *TreeP* is that each virtual node can be elected among real nodes by its performance based on its characteristics such as power, network capacity, connection bandwidth, storage capacity, etc. More details about *TreeP* and its performance can be found in [41].

In order to deploy our *Optimal kdContruct* algorithm on a high performance computing platforms (centralised or distributed), we firstly construct the *TreeP* based on the availability of resources (processing units) so that each node of *TreeP* is a process. These processes are located in one or different computing nodes across the network. Note that each node of our *TreeP* in this case only has two children (left and right). Next, we construct our *kd-tree* on *TreeP* from the root level i.e. the root node of *kd-tree* is built at the root node of *TreeP*. Now we can apply *Algorithm 3* below to build our *kd-tree*. In this algorithm, *TpNode* is a node of *TreeP* tree.

*Algorithm 3 Par-Optimal kdContruct*
1: **procedure** kdConstruct (*trainingSet, TpNode*)
2: **if** *trainingSet.size*() <= Leafsize **then**
3: **return** *kdtree.leafnode*      // Returns a kdTree
4: **else**
5: (*s, val*) ← chooseSplit(*trainingSet*) // *s* is splitting dimension, chooseSplit function based on sliding midpoint rudes
6: *trainLeft* ← *{x ∈ trainingSet : $x_s$ < val }*
7: *trainRight* ← *{x ∈ trainingSet : $x_s$ ≥ val }*
8: *kdLeft* ← kdConstruct(*trainLeft, TpNode.left*)
9: *kdRight* ← kdConstruct(*trainRight, TpNote.right*)
10: **return** *kdtree(s, val, kdLeft, kdRight)*
11: **end procedure**

By using the efficient *TreeP* topology, we can deploy our new approach on high performance computing platforms. We can moreover implement the brute-force algorithm (*Algorithm 2*) on this topology to improve its performance in terms of running time with the complexity of *O(logn)*.

Besides, we also develop a tool that implements our algorithms and allows us to process the LiDAR data called LiDAR Plotter (Fig.5). Our tool has the important functions such as displaying points loaded from LAS file in 2D and 3D mode; supporting basic view interactions: moves, rotates, zooms; supporting cropping by specified polygon to reduce the displaying area; displaying the information about LAS file, etc.

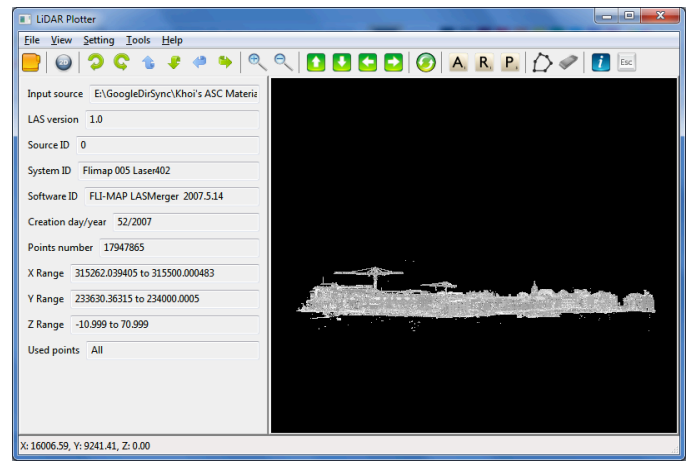

Fig.5 LiDAR Plotter

V. CONCLUSION AND FUTURE WORK

LiDAR data represent the basis for some of the fastest growing datasets from both density and availability perspectives. Today, aerial imagery and aerial laser scanning are nearly commonplace for general mapping, three-dimensional city modelling, and disaster management. Despite this data explosion, there has yet to be a mechanism to process such information in an efficient manner, to say nothing of doing so via the Internet. Processing is critical not simply for visualisation, but for the merging and querying of multiple datasets and the subsequent processing for segmentation, as well as computational modelling. In this paper, we conduct comparative studies of existing libraries and methods to determine the key issues that affect the performance of LiDAR processing. We also propose a new strategy for ALS data processing. We describe moreover the ability of improving the performance of our approach by integrating parallel computing based on an efficient network topology *TreeP*.

Experimental results of parallel approach for both *kd-tree* construction and brute-force searching with very large size of LiDAR data are also being produced. These results will allow us to test and evaluate the robustness of our approach.

## REFERENCES

[1] E. Naesset, "Determination of mean tree height of forest stands using airborne laser scanner data," *ISPRS Journal of Photogrammetry and Remote Sensing* vol. 52, no. 2, pp. 49-56, 1999.
[2] M. Kada, and L. McKinley, "3D building reconstruction from LiDAR based on a cell decomposition approach," *International Archives of Photogrammetry, Remote Sensing and Spatial Information Sciences* vol. 38, no. Part 3/W4, 2009.
[3] A. Boyko, and T. Funkhouser, "Extracting roads from dense point clouds in large scale urban environment," *ISPRS Journal of Photogrammetry and Remote Sensing,* vol. 66, no. 6, pp. S2-S12, 2011.
[4] D. F. Laefer, and A. R. Pradhan, "Evacuation route selection based on tree-based hazards using light detection and ranging and GIS," *Journal of transportation engineering,* vol. 132, no. 4, pp. 312-320, 2006.
[5] Truong-Hong L, Laefer DF, Hinks T *et al.*, "Flying Voxel Method with Delaunay Triangulation Criterion for Façade/Feature Detection for Computation," *ASCE Journal of Computing in Civil Engineering,* vol. 26, no. 6, pp. 691–707, 2012.
[6] L. geosystems. "Leica ALS80 Airborne Laser Scanner," http://www.leica-geosystems.com/en/Leica-ALS80-Airborne-Laser-Scanner_105650.htm.
[7] D. L. Laefer, C. O'Sullivan, H. Carr *et al.*, "Aerial laser scanning (ALS) data collected over an area of around 1 square km in Dublin city in 2007," UCD Library, University College Dublin, 2014.
[8] A. H. Nederland. "Actualisatie van het 2," http://www.ahn.nl/index.html.
[9] R. Swart, "How to handle the Up-to-date Height Model of the Netherlands: detailed, precise, but so huge!," in Management of massive point cloud data: wet and dry, Oracle, De Meern, The Netherlands, 2009.
[10] H. Wu, X. Guan, and J. Gong, "ParaStream: A parallel streaming Delaunay triangulation algorithm for LiDAR points on multicore architectures," *Computers & Geosciences* vol. 37, no. 9, pp. 1355-1363, 2011.
[11] B. Yang, and Z. Dong, "A shape-based segmentation method for mobile laser scanning point clouds," *ISPRS Journal of Photogrammetry and Remote Sensing,* vol. 81, pp. 19-30, 2013.
[12] J. Elseberg, D. Borrmann, and A. Nuchter, "Efficient processing of large 3d point clouds." pp. 1-7.
[13] S. H. Han, J. Heo, H. G. Sohn *et al.*, "Parallel processing method for airborne laser scanning data using a PC cluster and a virtual grid," *Sensors,* vol. 9, no. 4, pp. 2555-2573, 2009.
[14] M. Isenburg, Y. Liu, J. Shewchuk *et al.*, "Streaming computation of Delaunay triangulations." pp. 1049-1056.
[15] J. Bedkowski, K. Majek, and A. Nüchter, "General purpose computing on graphics processing units for robotic applications," *Journal of Software Engineering for Robotics,* vol. 4, no. 1, pp. 23-33, 2013.
[16] M. Liu, F. Pomerleau, F. Colas *et al.*, "Normal estimation for pointcloud using gpu based sparse tensor voting." pp. 91-96.
[17] G. Sithole, and G. Vosselman, "Experimental comparison of filter algorithms for bare-Earth extraction from airborne laser scanning point clouds," *ISPRS Journal of Photogrammetry and Remote Sensing,* vol. 59, no. 1, pp. 85-101, 2004.
[18] S. Filin, and N. Pfeifer, "Segmentation of airborne laser scanning data using a slope adaptive neighborhood," *ISPRS Journal of Photogrammetry and Remote Sensing,* vol. 60, no. 2, pp. 71-80, 2006.
[19] C.-K. Wang, and Y.-H. Tseng, *Dem generation from airborne lidar data by an adaptive dual-directional slope filter*: na, 2010.
[20] D. Tóvári, and N. Pfeifer, "Segmentation based robust interpolation-a new approach to laser data filtering," *IAPRS,* vol. 36, no. 3, pp. W19, 2005.
[21] P. Axelsson, "DEM generation from laser scanner data using adaptive TIN models," *International Archives of Photogrammetry and Remote Sensing,* vol. 33, no. B4/1; PART 4, pp. 111-118, 2000.
[22] K. Kraus, and N. Pfeifer, "Determination of terrain models in wooded areas with airborne laser scanner data," *ISPRS Journal of Photogrammetry and Remote Sensing,* vol. 53, no. 4, pp. 193-203, 1998.
[23] G. Sohn, and I. Dowman, "Terrain surface reconstruction by the use of tetrahedron model with the MDL criterion," *International Archives of Photogrammetry Remote Sensing and Spatial Information Sciences,* vol. 34, no. 3/A, pp. 336-344, 2002.
[24] K. Zhang, S.-C. Chen, D. Whitman *et al.*, "A progressive morphological filter for removing nonground measurements from airborne LIDAR data," *Geoscience and Remote Sensing, IEEE Transactions on,* vol. 41, no. 4, pp. 872-882, 2003.
[25] G. V. Vosselman, and H.-G. Maas, *Airborne and terrestrial laser scanning*: Whittles, 2010.
[26] R. Schnabel, R. Wahl, and R. Klein, "Efficient RANSAC for PointCloud Shape Detection." pp. 214-226.
[27] G. Vosselman, and S. Dijkman, "3D building model reconstruction from point clouds and ground plans," *International Archives of Photogrammetry Remote Sensing and Spatial Information Sciences,* vol. 34, no. 3/W4, pp. 37-44, 2001.
[28] B. Gorte, "Segmentation of TIN-structured surface models," *International Archives of Photogrammetry Remote Sensing and Spatial Information Sciences,* vol. 34, no. 4, pp. 465-469, 2002.
[29] S. Filin, "Surface clustering from airborne laser scanning data," *International Archives of Photogrammetry Remote Sensing and Spatial Information Sciences,* vol. 34, no. 3/A, pp. 119-124, 2002.
[30] A. D. Hofmann, H.-G. Maas, and A. Streilein, "Derivation of roof types by cluster analysis in parameter spaces of airborne laserscanner point clouds," *IAPRS International Archives of Photogrammetry and Remote Sensing and Spatial Information Sciences,* vol. 34, no. Part 3, pp. W13, 2003.
[31] J. H. Friedman, J. L. Bentley, and R. A. Finkel, "An algorithm for finding best matches in logarithmic expected time," *ACM Transactions on Mathematical Software (TOMS),* vol. 3, no. 3, pp. 209-226, 1977.
[32] https://github.com/grantbrown/laspy
[33] http://www.pyzo.org/whypython.html
[34] NumPy, http://www.numpy.org/
[35] SciPy, http://docs.scipy.org/doc/scipy/reference/tutorial/general.html
[36] C. Bauckhage, *NumPy / SciPy Recipes for Data Science: Squared Euclidean Distance Matrices*, ResearchGate, 2014
[37] http://docs.scipy.org/doc/numpy/reference/generated/numpy.vectorize.html
[38] http://stackoverflow.com/questions/22581763/python-numpy-apply-a-function-to-each-row-of-a-ndarray
[39] J. L. Bentley, Multidimensional binary search trees used for associative searching, Communications of the ACM , 1975
[40] S. Maneewongvatana and D. M. Mount, It's okay to b e skinny, if your friends are fat, 4th Annual CGC Workshop on Computational Geometry, 1999
[41] E. EDI, M-T. Kechadi, and R. McNulty. TreeP: A Self- Reconfigurable Topology for Unstructured P2P Systems. LNCS on State-of-the-Art in Scientific & Parallel Computing, Vol. 4699 p.1136-1146, 2007
[42] Comer D. Ubiquitous B-tree ACM Computing Survey, Vol.11, No.2: 121-137
[43] X. Guan and H. Wu, "Leveraging the power of multi-core platforms for large-scale geospatial data processing: Exemplified by generating DEM from massive LiDAR point clouds," Computers & Geosciences, no. 36, pp. 1276-1282, 2010.

[44] D. Liao and S. Y. Berkovich, "A Multi-Core Pipelined Architecture for Parallel Computing," Parallel & Cloud Computing, vol. Vol. 2, no. Iss. 2, pp. 49-57, 2013.
[45] X. L. a. Y. Z. Xiangyun Hu, "Fast Filtering of LiDAR Point Cloud in Urban Areas Based on Scan Line Segmentation and GPU Acceleration," IEEE GEOSCIENCE AND REMOTE SENSING LETTERS, vol. 10, no. 2, pp. 308-312, 2013.
[46] K. Shih, A. Balachandran, K. Nagarajan, B. Holland, C. Slatton and A. George, "Fast Real-Time LIDAR Processing on FPGAs," in THE 2008 INTERNATIONAL CONFERENCE ON ENGINEERING OF RECONFIGURABLE SYSTEMS & ALGORITHMS, Las Vegas, 2008.
[47] S. Maneewongvatana and D. M. Mount, Analysis of Approximate Nearest Neighbor Searching with Clustered Point Sets, Data Structures, Near Neighbor Searches, and Methodology, 2002.
[48] J. Elseberg, S. Magnenat, R. Siegwart and A. Nuchter, Comparison of nearest-neighbor-search strategies and implementations for efficient shape registration, Journal of Software Engineering for Robotics, 2012.